# Adsorption configurations of Co-phthalocyanine on $In_2O_3(111)$

M. Wagner[1*], F. Calcinelli[2], A. Jeindl[2], M. Schmid[1], O.T. Hofmann[2], U. Diebold[1]

[1] Institute of Applied Physics, TU Wien, 1040 Vienna, Austria
[2] Institute of Solid State Physics, Graz University of Technology, 8010 Graz, Austria

*corresponding author: wagner@iap.tuwien.ac.at

**Abstract**

Indium oxide offers optical transparency paired with electric conductivity, a combination required in many optoelectronic applications. The most-stable $In_2O_3(111)$ surface has a large unit cell (1.43 nm lattice constant). It contains a mixture of both bulk-like and undercoordinated O and In atoms and provides an ideal playground to explore the interaction of surfaces with organic molecules of similar size as the unit cell. Non-contact atomic force microscopy (nc-AFM), scanning tunneling microscopy (STM), and density functional theory (DFT) were used to study the adsorption of Co-phthalocyanine (CoPc) on $In_2O_3(111)$. Isolated CoPc molecules adsorb at two adsorption sites in a 7:3 ratio. The Co atom sits either on top of a surface oxygen ('F configuration') or indium atom ('S configuration'). This subtle change in adsorption site induces bending of the molecules, which is reflected in their electronic structure. According to DFT the lowest unoccupied molecular orbital of the undistorted gas-phase CoPc remains mostly unaffected in the F configuration but is filled by one electron in S configuration. At coverages up to one CoPc molecule per substrate unit cell, a mixture of domains with molecules in F and S configuration are found. Molecules at F sites first condense into a F-(2×2) structure and finally rearrange into a F-(1×1) symmetry with partially overlapping molecules, while S-sited molecules only assume a S-(1×1) superstructure.

Atomic force microscopy; Density functional theory; Organic molecule on $In_2O_3(111)$; Electronic structure of organic molecules

# 1. Introduction

Many metal oxide surfaces feature large unit cells composed of differently (under-) coordinated cations and anions, making the unit cell inhomogeneous in topography, electronic properties, and reactivity. Therefore, the properties of adsorbed molecules are expected to strongly depend on the adsorption site. We demonstrate this by using $In_2O_3(111)$ as substrate. This surface is bulk-terminated but exhibits various combinations of both bulk-like and undercoordinated cations and oxygen atoms in its unit cell. We study the adsorption of cobalt-phthalocyanine, an organic molecule with a size comparable to the $In_2O_3(111)$ unit cell.

Indium oxide is a technologically relevant transparent conductive oxide (TCO), commonly doped with Sn (then called ITO) to increase its intrinsic conductivity. It is utilized as the transparent top electrode in optoelectronic devices. When the smaller $Sn^{4+}$ replaces some $In^{3+}$, the lattice is slightly compressed [1]. However, the structure of the (111) surface is essentially identical to the undoped material [2]. This makes undoped $In_2O_3(111)$ a good model substrate to study surface phenomena such as the adsorption of organic molecules. $In_2O_3$ crystallizes in a cubic lattice with the bixbyite structure. Its (111) surface has a remarkably large lattice parameter of $a = 1.431$ nm, which gives rise to an inhomogeneous unit cell with 3-fold symmetry. At the surface, 40 atoms are arranged in an O–In–O trilayer leading to several combinations of 6- and 5-fold coordinated In with 4- and 3-fold coordinated O atoms. The atomic structure of the surface is described in detail in the Supplementary Material and Ref. [3].

Phthalocyanines (Pc) are synthetic molecules consisting of four isoindole (benzo-fused pyrrole) units linked by N atoms. The molecules feature a fully conjugated aromatic system, hence a strong absorption in the visible range of light, and good chemical and thermal stability. The free base (2HPc) can be modified by substituting the protons in the central cavity by a metal atom (MPc). MPcs are commercially utilized as dye and pigment. Combining metalation with modifications by functional side groups makes this p-type semiconductor attractive in photovoltaics [4, 5] and catalysis [6]. Surface science investigations of (M)Pc on crystalline metal and metal oxide substrates have provided rich information about structural and electronic details of single molecules [7-11], self-assembly into layers [12-19], and thin film growth behavior [20-22]. Another hot topic in this field are single-molecule magnets where the Kondo temperature of MPc is related to the interaction with a metallic substrate [23-27]. An overview on phthalocyanines in surface science is given in Ref. [28].

In this work, we report on the adsorption of Co-phthalocyanine (CoPc) molecules on $In_2O_3$(111), employing low-temperature scanning tunneling microscopy (STM) and non-contact atomic force microscopy (AFM) as well as density functional theory (DFT). We find that CoPc coexists in two different adsorption configurations and that the interaction with the substrate influences the geometry and distortion of the molecule.

## 2. Methods

The experiments were conducted with an Omicron LT-STM/AFM ($T$ = 4.8 K, base pressure $5\times10^{-11}$ mbar) equipped with qPlus sensors [29] and a differential preamplifier [30], and a separate wire for the tunneling current [31]. Electrochemically etched W tips (20 µm diameter) were glued to the free prong of the tuning fork. The parameters of the qPlus sensor were: Resonance frequency $f_R$ = 68.5 kHz, spring constant $k \approx$ 5,400 N/m, and quality factor $Q \approx$ 96,000. The tip was prepared on $In_2O_3$(111) by applying voltage pulses and dipping it gently into the surface until a frequency shift less than −3 Hz was reached at the setpoint of +1 V and 20 pA. The (111) surface of nominally undoped indium oxide single crystals was prepared by sputtering and annealing in oxygen; the preparation recipe is described elsewhere [3]. Cobalt(II) phthalocyanine (Sigma-Aldrich, β-form, dye content 97 %), was evaporated by thermal sublimation (~490 °C, Omnivac four-pocket evaporator). During the deposition, the sample was kept at room temperature. Afterwards it was annealed at 200 °C to desorb co-adsorbed water. The size of phthalocyanine (~1.5 nm diameter including H atoms, measured across opposite isoindole units, area ~1.13 nm$^2$) is comparable to the size of the $In_2O_3$(111) unit cell (1.77 nm$^2$). A (1×1) superstructure consisting of 1 CoPc molecule per substrate unit cell is thus possible; we refer to such a coverage as one monolayer (ML) CoPc.

The computational results were obtained using DFT, employing the PBE exchange correlation functional [32] with the MBD-NL as dispersion correction [33] and the default tight numerical atomic orbitals provided within the FHI-aims package [34]. The $In_2O_3$(111) periodic slab was simulated with four $O_{12}$-$In_{16}$-$O_{12}$ trilayers, corresponding to a thickness of 1.1 nm. The adsorbates were placed only on only one side of the slab. The repeated slab approach with a dipole correction [35] was utilized to electrostatically decouple periodic replicas normal to the surface. The geometry optimizations were performed in a ($\sqrt{3} \times \sqrt{3}$)R30° cell (i.e., three times the surface area of the primitive $In_2O_3$(111) cell). The experimental results were used as the initial guess and the geometries were optimized until the maximum remaining force dropped below 0.01 eV/Å for all atoms in the unit cell. During the geometry optimization, we employed a Gamma-centered k-grid with 2×2×1 k-points. The evaluation of the electronic structure was performed with a denser 4×4×1 k-grid. All calculations were performed spin-unrestricted with a broadening of 0.03 eV and a SCF convergence threshold of $1\times10^{-5}$ eV.

## 3. Results and Discussion

### 3.1. Isolated CoPc molecules

After the deposition of ~0.15 ML of CoPc on $In_2O_3$(111) and heating to 200 °C, STM reveals single molecules and pairs of molecules. All adsorption configurations are present in three symmetry-equivalent orientations related to the 3-fold symmetry of the substrate. The molecules are randomly distributed across the terraces without any preference for adsorption at step edges or defects. Fig. 1 shows STM images acquired at different bias voltages: Panel (a) probes the energy region between the $In_2O_3$ Fermi energy and the LUMO of the CoPc molecules; panel (b) shows the same surface area as in (a) at the energy where the LUMO of MPc on metal surfaces is typically found. Two different appearances of CoPc are identified, labeled F (referring to the flowerlike appearance in Fig. 1(b)) and S (referring to the second, 'square' appearance). As will be shown in the following, these are related to different adsorption sites and geometries of the molecules causing different electronic properties. Species F, the majority of the molecules (>70 %), appears as is typical for STM of Co-phthalocyanine on any substrate [36-38]. Within the HOMO-LUMO gap, the molecule resembles a cross with a bright center (panel (a)); when imaging the LUMO, a bright protrusion is located at the Co atom with less intense double features at each benzene (panel (b)). The minority species S (<30 %) appears as a bright cross with a dark rim around the Co atom; this appearance remains unchanged for bias voltages of +(0.3–1.0) V. At higher STM bias voltages, the nodal structure of the LUMO starts to appear (see Fig. 1(c)), indicating that the LUMO of S-sited CoPc is located at higher energies compared to CoPc in F site. At these high STM bias voltages (panel c), the CoPc in F and S site resemble each other, except that the center of the S molecule remains dark.

During the STM image acquisition and while tunneling into the LUMO, the STM tip occasionally induced changes of some molecules from S to F and *vice versa*, see the molecules inside the white dashed frame of Figs. 1(a, b). This confirms that both configurations are indeed CoPc and the minority species S is not the de-metalated 2HPc molecule. The presence of the tip influenced only a few molecules; most molecules were immune to switching.

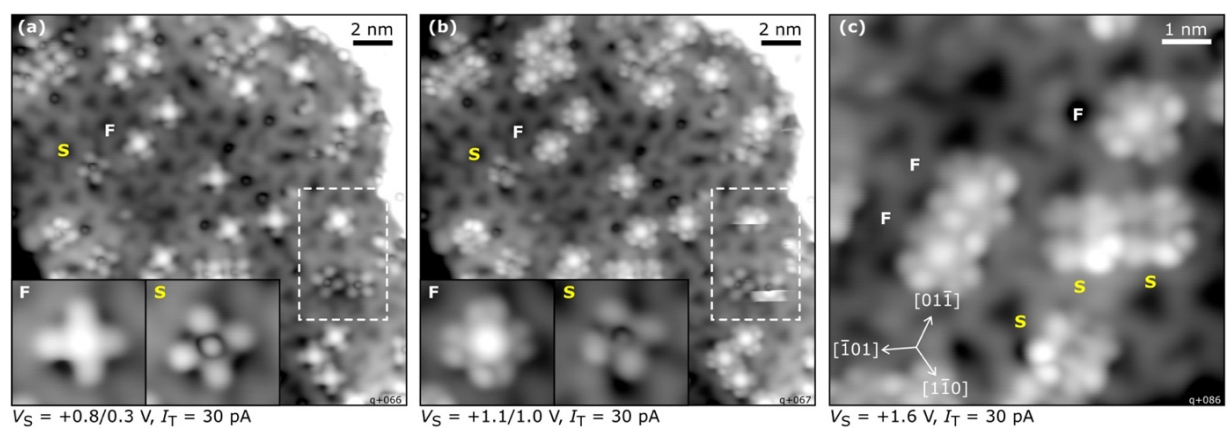

Fig. 1. Adsorption of ~0.15 ML CoPc molecules on $In_2O_3$(111). Constant-current STM images are acquired (a) within the HOMO-LUMO gap of the molecules and (b) at the LUMO energy of the F type CoPc with the tunneling parameters indicated at the bottom of each panel separately for main image/inset). Two different adsorption configurations are apparent, labeled F (flower) and S (square). Tip-induced F ↔ S conversion of individual molecules during image acquisition was occasionally observed (dashed white frame). (c) Appearance of both configurations at higher sample bias voltage.

The adsorption configuration was further analyzed with constant-height nc-AFM and DFT calculations. Fig. 2(a) shows an STM image of two differently oriented CoPc molecules in the F site; the inset displays the flower-like appearance characteristic for STM at voltages corresponding to the LUMO. Imaging the same molecules with constant-height nc-AFM, panel (b), reveals details of their adsorption geometry. Three isoindole units of the molecule are easily identified by their terminating benzene rings, including the –N= link that connects the isoindoles. The fourth benzene is less pronounced, indicating that it is located closer to the surface, i.e., bent downwards. The Co atom in the center of the CoPc molecule is attractive to the tip and appears dark. The long axis of the molecule (dashed black line), defined by the isoindole bent towards the surface and the opposite, bright one, is always aligned with ⟨1$\bar{1}$0⟩ directions. The upper molecule eventually changed its configuration from F to S when imaging at +1.0 V, clearly distinguishable by its modified appearance in Fig. 2(c). In AFM the S molecule also clearly differs from its previous F-site appearance (panel (d)). Only one benzene is clearly visible, while the rest of the molecule is rather featureless, suggesting that the molecule in the S site is distorted differently and one isoindole group is pointing away from the surface. The molecule in F site (bottom left of panel (d)) now appears less pronounced compared to panel (b), but this is only due to the acquisition heights: the upwards-tilted benzene in S site requires the AFM tip to stay further away from the surface.

The adsorption conformations, deduced from STM and AFM data and refined by DFT calculations, are presented in Figs. 2(e–g). The AFM image of Fig. 2(d) already revealed that the molecules in F and S sites are arranged differently, and the computational results confirmed this. In the F site, one isoindole group is bent towards the surface while the rest of the molecule is almost parallel to the surface and only slightly distorted, see the side view of the adsorption configuration in Fig. 2, panel (e). The S configuration exhibits a stronger distortion of the whole molecule towards the surface and an upwards tilt of one isoindole unit. This up-bending is more pronounced than that of the others units in the S site or any isoindole in the F configuration (panel (g)). This causes the high contrast of one benzene ring compared with the low contrast of the rest of the molecule in AFM (Fig. 2(d)). The higher adsorption energy of the F conformation (–3.21 eV) compared to the S conformation (–2.84 eV) is consistent with the observed preponderance for F.

Switching from F to S requires the molecule to rotate anticlockwise (indicated by the dashed black and yellow lines in Fig. 2(a–d)) and shift laterally. This moves the Co atom from its position on top of an O($\gamma$) to the neighboring In(c) atom see Ref. [3] and the supplementary material for the designations of the O and In sites. In Fig. 3(f, h), these substrate atoms are labeled by white circles where they are not covered by the Co atom. (Thus, each white circle indicates the Co position in the other configuration.)

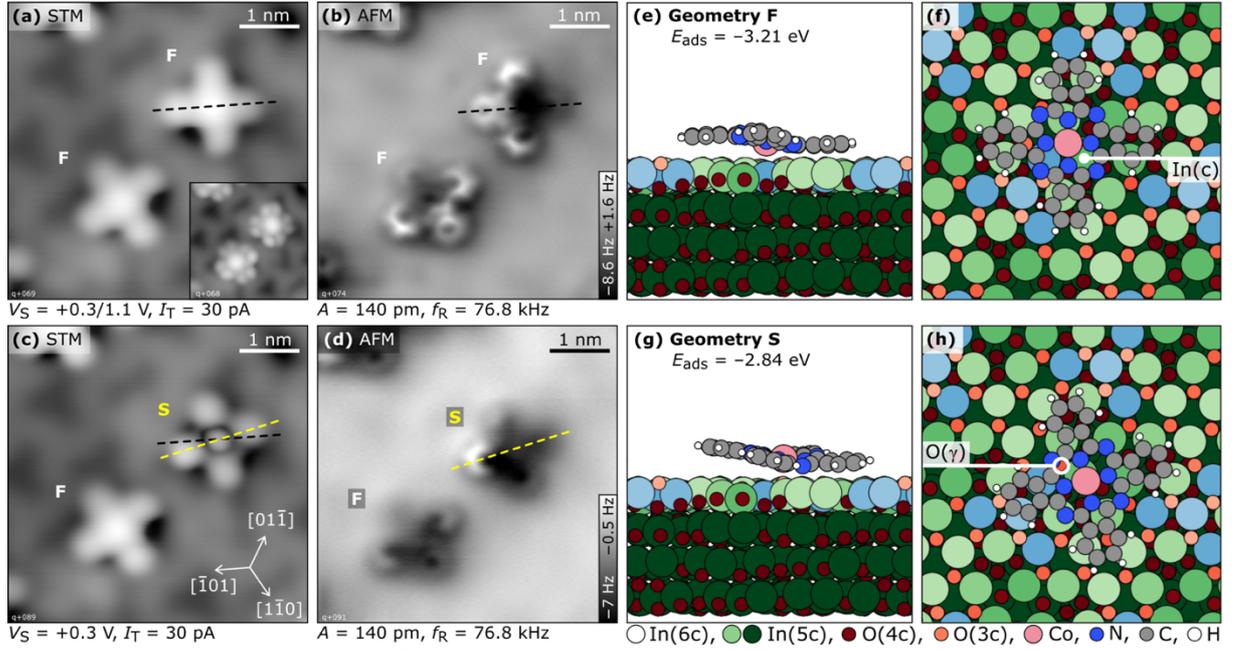

Fig. 2. The two adsorption configurations of CoPc on $In_2O_3(111)$. (a, b) STM and constant-height AFM images of two molecules in symmetry-equivalent rotational orientations in F sites. The long axis of the upper molecule is indicated by a black dashed line. (c, d) Comparison of CoPc in F and S sites. The upper molecule switched from the F to the S configuration during STM imaging (not shown here), which requires a rotation (indicated by a yellow dashed line) and a small translation of 2.2 Å. The AFM image in (d) was acquired at a larger tip height than (b). (e–h) Geometry-optimized DFT results for the F and S adsorption in side (e, g) and top views (f, h). Since the adsorption site of the Co atom is hidden in the top views, these positions are indicated in (f, h) for the respective other configuration. Substrate atoms in the top layer are colored according to their z coordinate, with lower and higher-lying atoms in darker and lighter color, respectively. The 6-fold-coordinated In atoms are colored in blue, 5-fold-coordinated In atoms in green, and O atoms in red.

To gain further insight into the difference between the substrate-adsorbate interactions of the two configurations, we performed projections of the total density of states of the adsorbed molecule onto the molecular orbitals of the free molecule (MODOS). All energetic positions mentioned in the following are given relative to the valence band maximum of $In_2O_3$. We focus our attention on the frontier orbitals of the different configurations, using the undisturbed gas-phase molecule as reference. Due to the spin-polarization of Co, the orbitals are singly occupied (the degeneracy within one spin channel notwithstanding). This is in contrast to the usual doubly (un)occupied HOMO (LUMO); we therefore refer to the frontier orbitals H-SOMO (highest singly occupied molecular orbital) and L-SUMO (lowest singly occupied molecular orbital). The H-SOMO is centered at the ligands, while the L-SUMO is located mainly on the Co atom. Disturbing the geometry of the molecule into the F or S configuration does not change the energetic ranking or form of the orbitals (see Fig. S2). Upon adsorption of the CoPc in F configuration on the surface, the MODOS shows that the ligand-centered H-SOMO stays occupied at +0.7 eV (blue curve in Fig. 3(a)) while the metal-centered L-SUMO remains empty at +1.8 eV (orange curve in Fig 3(a)). However, the broad shape of the L-SUMO indicates that it strongly hybridizes with the substrate. For the molecule in S configuration, adsorption on the surface has a substantial influence on the electronic structure. The H-SOMO moves slightly down in energy compared to the F configuration (to +0.6 eV). The metal-centered orbital on the other hand now strongly hybridizes with the surface, appearing as a peak around +0.25 eV and a second smaller peak around +0.9 eV (see orange curve in Fig. 3(b)). Thus, the former L-SUMO, which stays unoccupied in the F configuration, is now found mostly under the Fermi energy and contributes to the bonding between the molecules and the surface. Besides the Co-centered orbital, also several other orbitals of the S molecule shift with respect to the F configuration (see Supporting Information). Especially notable is the emergence of the molecular peak at VBM+0.9 eV, which can be partially attributed to the L-SUMO and partially to a lower lying π orbital (see Figure S3 for details).

Overall, the DFT analysis thus indicates that the two configurations interact quite differently with the surface. The F-configuration closely resembles the situation in the gas phase. The S-configuration, although energetically less favorable, binds through hybridization of the frontier orbitals with the surface. The reason for the lower adsorption energy of geometry S is threefold. First, the adsorption onto the S configuration entails a larger Pauli-repulsion due to the overall closer adsorption distance. The energetic cost for the deformation of the molecule is also increased. While the geometric distortion into the F geometry costs 0.23 eV (compared to the free gas phase CoPc), it costs 0.32 eV for the S geometry. Third, the contribution of the van der Waals energy to the adsorption energy for S is lower compared to F (–2.59 eV versus –2.70 eV) due to one isoindole unit being bent further away from the substrate. The electronic structure projected onto all frontier orbitals (including those in Fig. 3) for both configurations is shown in the Supplementary Information.

The metal-centered state can be used as a measure for the interaction with the substrate, as the Co $d_{z^2}$-type orbital protrudes strongly above and below the molecular backbone. The occupation of this orbital in configuration S, in contrast to both the gas phase and F configuration, where it remains unoccupied, suggests a strong interaction with the substrate, which, however, comes at the cost of a more considerable distortion.

The large difference in electronic structure explains the different appearance of the two molecules in empty-states STM images.

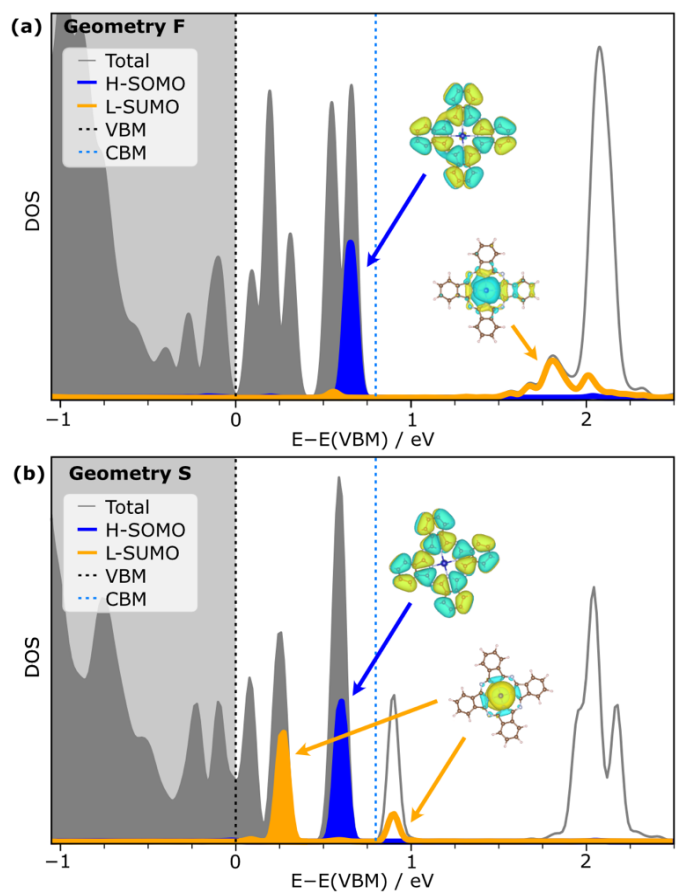

Fig. 3. Electronic DOS of CoPc adsorbed in (a) F and (b) S geometry. The maximum of the substrate valence band is taken as 0, and the area occupied by the valence band is shaded in light grey. Filled and empty states are represented by filled and empty curves, respectively. Projection of the total density of states on two selected orbitals of the undistorted CoPc: The ligand-centered H-SOMO (blue), and the metal-centered L-SUMO (orange).

*3.2. CoPc coverages up to the monolayer limit*

When the coverage is increased, the molecules start to order. Fig. 4(a-d) shows the arrangement in pairs, where CoPc molecules in F or S sites occupy neighboring unit cells without molecular overlap. Due to the different adsorption sites, and, hence, different rotational orientation of the molecules, the alignment of the molecules within the chains differs, even though both types of chains run along the $\langle 1\bar{1}0 \rangle$-type directions of the substrate.

At CoPc coverages below ~0.5 ML (i.e., 0.5 molecules per $In_2O_3(111)$ unit cell), the F:S ratio stays constant and the formation of pairs and short one-dimensional chains follows the trend already started in Fig. 5. Within a chain, the molecules assume the same orientation to avoid overlap and, consequently, the S and F site do not mix. Chains made from F-site molecules often change their direction by 120° (Fig. 5(e, f)), which is possible without molecular overlap while keeping the same orientation of the molecule, see light blue molecule in Fig. 5(a). The only difference between these differently oriented chains of the same molecular orientation is the surface structure in-between adjacent molecules.

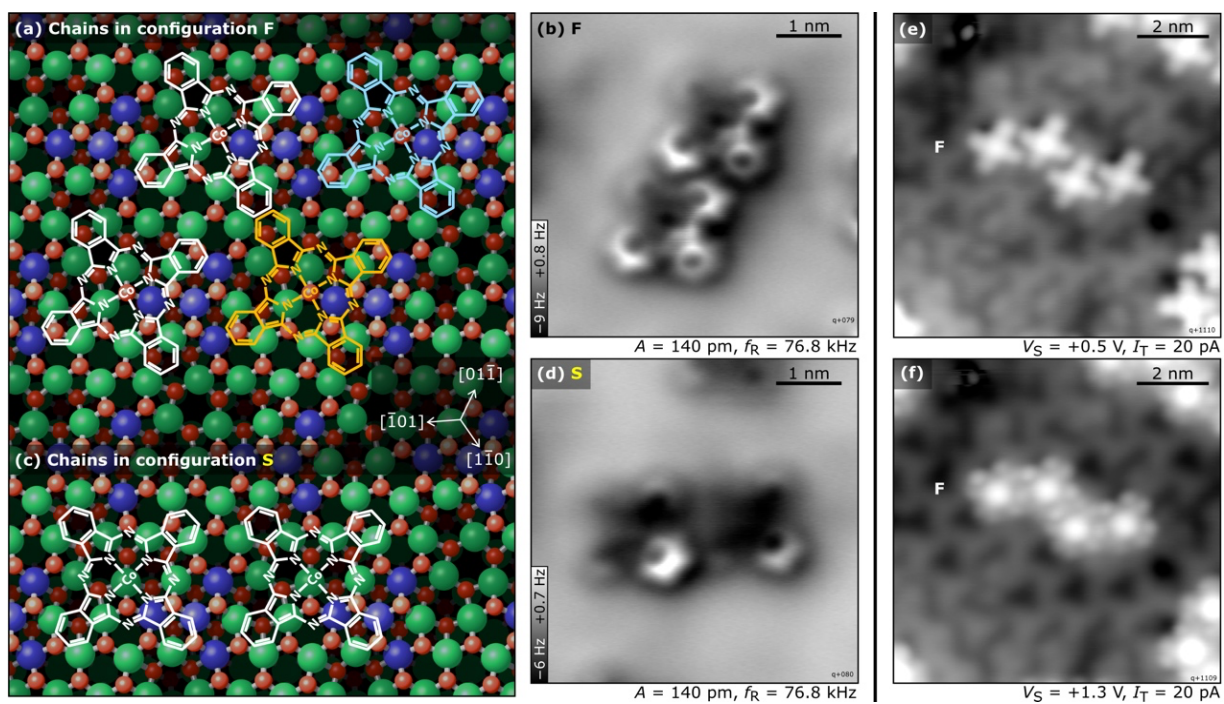

Fig. 4. Pairs formed by CoPc in F (a, b) and S (c, d) conformation. The schematics show the CoPc gas-phase structure superimposed onto the atomic model of the $In_2O_3$(111) surface, as derived from the corresponding AFM images. The orange molecule in (a) visualizes the steric hindrance of flat F molecules in an (1×1) structure. (e, f) STM images of a zigzag chain consisting of molecules in F site acquired at bias voltages in the HOMO-LUMO gap (e) and at the LUMO energy (f).

After further increasing the CoPc coverage to ~0.75 ML, see Fig. 5(a), both F and S molecules form small domains of ordered structures, and the surface appears fully covered. The images in Fig. 5 were taken at STM bias voltages corresponding to the LUMO, where S molecules are imaged as four small bright blobs surrounding a dark center, and F molecules are imaged as large protrusion surrounded by eight smaller protrusions. Molecules in S sites maintain their assembly into chains, sometimes forming small two-dimensional domains (Fig. 5(b)) with S-(1×1) symmetry, even though the majority stays as single molecules or pairs. In contrast to the molecules in S sites, there are no areas of molecules in F sites with a local coverage of 1 ML in a (1 × 1) arrangement. There are still short chains along one of the ⟨1$\bar{1}$0⟩ directions, but in the direction perpendicular to these chains the distance to the nearest molecules is always larger than what it would be in a (1 × 1) structure. This is explicable since the kink structure in Fig. 4(a) and Fig. 4(e, f) cannot be extended to full monolayer coverage due to steric hindrance (orange molecule in Fig. 4(a)). A statistical analysis of Fig. 5(a) reveals that the probability for an F molecule to have an F neighbor with the same rotational orientation at a distance of the substrate lattice constant $a$ is only about 4 %, while it is 30 % in a distance of $2a$. It is also more likely for the nearest neighbors of an F molecule to have one of the two other orientations than having the same orientation. In a few places, this results in a local (2 × 2) arrangement (one cell is marked by a white rhombus in Fig. 5(c)), but mostly in an apparently disordered pattern of the F molecules. A building block consisting of three molecules in different orientations is shown in Fig. 5(c); the individual molecules are marked by black crosses. In AFM (panel (d)) molecules in both, F and S conformations look identical to the single molecules of Fig. 2(d). This confirms that all molecules retain the adsorption sites and geometry described in section 3.1.

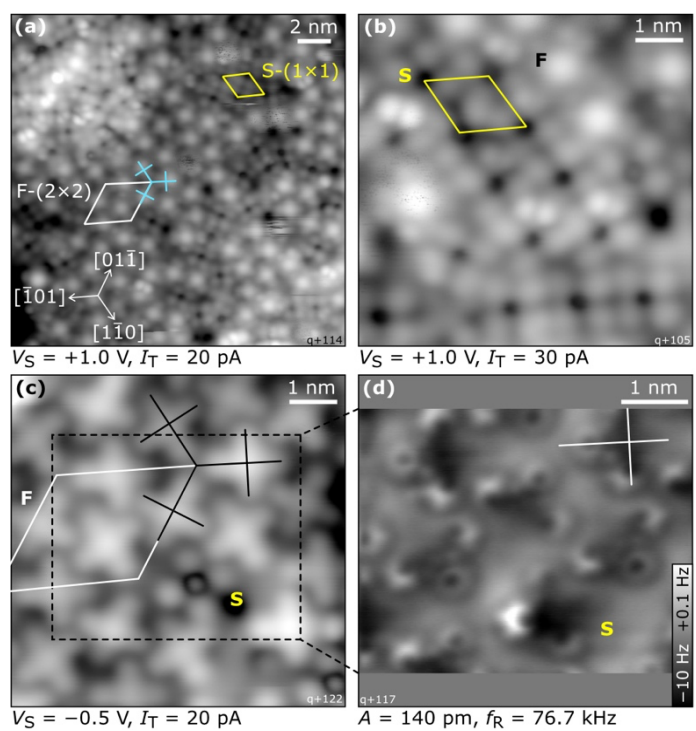

Fig. 5. Structural mixture at a coverage of ~0.75 ML CoPc on $In_2O_3$(111). (a) Overview, showing a (2×2) cell formed by molecules in F site. CoPc molecules in S sites are present as single molecules embedded in F areas or as short chains; neighboring chains form a (1×1) structure. (b) STM image of a small patch with S-(1×1) structure. (c) Molecular arrangement of a F-(2×2) cell. Black crosses indicate the three molecules forming the building block. (d) Constant-height AFM image of the F-(2×2) molecules with a single molecule in an S site (black dashed frame in panel (c)).

Increasing the coverage to 1 ML of CoPc (i.e., to one molecule per $In_2O_3$(111) unit cell), see Fig. 6, leads to a rearrangement of the F-sited molecules into a (1×1) structure. This structure is present in three rotational domains. Since molecules in F site cannot arrange in (1×1) symmetry without overlapping (see Fig. 4(a)), they either change their adsorption site to S, where the different rotation allows (1×1) packing, or bend up on one side and overlap. Judging from the constant F:S ratio and the asymmetric appearance in STM (bright center and only one visible pair of protrusions, see Fig. 6(b)), it seems likely that one benzene ring of the molecules bends up and overlaps with the neighbor. Since there are no uncovered substrate areas any more, we cannot determine the adsorption sites of the molecules with respect to the substrate. Based on the relative positions of the S molecules in differently oriented domains we can verify that the S position remains unchanged, which is confirmed by the unchanged appearance of the molecules in the bias voltage range discussed above. The positions of the F molecules cannot be determined with high accuracy because of the lack of sharp features (like the dark center of the molecules in S sites) and the bending; nevertheless, we could determine that their positions remain the same within 0.3 nm; therefore, it is very likely that this adsorption site remains unchanged in spite of the bending and overlap at the periphery.

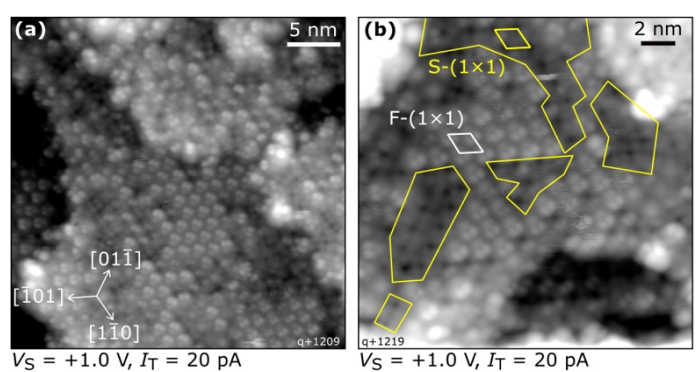

Fig. 6. Structural mixture at a coverage of ~1 ML CoPc on $In_2O_3$(111). Molecules in F site rearrange into a (1×1) structure with overlap between the molecules, causing them to bend up on one side (the double features correspond to the two maxima per benzene ring in Fig. 1(b)). Domains of the S-(1×1) structure are framed in yellow in panel (b).

## 4. Summary

We followed the structural evolution of CoPc on $In_2O_3$(111) up to the coverage of 1 molecule per substrate unit cell. For single molecules, two adsorption geometries were observed, where the Co atom of the molecule is located either on top of a particular surface O atom (F site, majority of molecules), or almost exactly on top of a specific In atom, with a different azimuthal orientation of the molecule (S site). In both sites, AFM experiments and DFT calculations reveal different but pronounced bending of the molecule. Comparing the projected DOS of the free CoPc molecule with those in F and S sites reveals that the F-site molecule is electronically similar to the free molecule. For CoPc in S site, however, the previously empty L-SUMO, centered at the Co atom, is now occupied. This strong interaction with the substrate leads to stronger distortions of the molecules, which result in a slightly lower adsorption energy of the S compared to the F site. Increasing the coverage stepwise leads to different arrangements of the F-site CoPc until a F-(1×1) structure with one molecule per substrate unit cell is formed. In contrast, the orientation of CoPc in S site allows a dense S-(1×1) packing without molecular overlap and no rearrangement is observed for the minority species.

**Acknowledgment**

This work was supported by the Austrian Science Fund (FWF), project V 773-N (Elise-Richter-Stelle, M.W.) and the project Y1157 (F.C., A.J., and O.T.H.). Computational results have been achieved using the Vienna Scientific Cluster (VSC). U.D. acknowledges funding from the European Research Council (ERC) under the European Union's Horizon 2020 research and innovation programme (grant agreement No. [883395], Advanced Research Grant 'WatFun').